# Photometric survey of the red giant V449 Cygni over half a century (1974 to 2022). A semi-regular variable star with a period of ~54 days.

Guy Boistel

GEOS, Groupe Européen d'Observation Stellaire (23 parc de Levesville, 28300 Bailleau l'Evêque, France)

**Abstract**:

This study sums up a major part of the red giant V449 Cygni measurements for the years 1974 through 2022, obtained by visual and CCD GEOS observers, the Unione Astrofili Italiani, and various automatized telescopes. It appears that the light variations of V449 Cyg in the years 1970-1990 corresponded more to an L-type star according to the GCVS classification, with irregular amplitude variations, sometimes marked by a quasi-periodicity of about 100 days. Since the mid-1990s, these variations have become more regular, with an amplitude of about 0.7 magnitude in V, and a period of about 54±1 days, as shown by various period search routines applied to the best recent CCD series. A longer period of about 2000 days is possible to describe the long term variation of V449 Cyg. Therefore, V449 Cyg appears to be an SRB-type star.

**Résumé**:

Cette étude revisite une grande partie des observations de la géante rouge V449 Cygni, observations visuelles et CCD obtenues par les observateurs du GEOS et de l'Unione Astrofili Italiani, ainsi que par divers télescopes automatiques, entre 1974 et 2022. Il apparaît que les variations de lumière de V449 Cyg correspondaient davantage dans les années 1970-1990 à une étoile de type L selon la classification du GCVS, avec des variations d'amplitude irrégulières, parfois marquées par une quasi-périodicité d'environ 100 jours. Depuis le milieu des années 1990, ces variations sont devenues plus régulières, d'une amplitude d'environ 0.7 magnitude en V, et d'une période d'environ 54±1 jours, comme le montrent divers périodogrammes appliqués aux meilleures séries CCD récentes. Une période plus longue d'environ 2000 jours est également possible. V449 Cyg apparaît dès lors comme une étoile de type SRB.

## 1. Introduction

**V449 Cygni** (GSC 2677.00291; J2000.0: α = 19h 53mn 20,97s; δ = +33° 57' 00,7''  (Skiff, 1999) is a red star of spectrum M1-M4 (Terrill, 1969). Its type of variation is LB: from GCVS (Samus et al., 2017), with a magnitude range of  7.4 – 9.07 (B filter). AAVSO-VSX server gives a magnitude range of 7.2-7.7 in V filter on the basis of TYCHO measurements. As an infrared source, V449 Cyg has also its IR reference (2MASS J19532097+3357006).

Recently, John R. Percy (2020) has published a long survey of data provided by ASAS-SN Sky Patrol database on some red stars. In his conclusion, Percy suggested new observations for some stars of his list, not observed by the ASAS-SN survey telescopes, including V449 Cyg, XY Lyr, FP Vir, IN Hya or OP Her (all these stars are part of the GEOS binocular priority research program). A few years earlier, Percy and Terziev (2011) even stated that V449 Cyg was not significantly variable!

Nevertheless, V449 Cyg is regularly observed by GEOS members since the beginning of the years 1970. Some annual visual light curves have been published by the GEOS in various papers (Figer, 1974; Boistel, 1984, 1987). Despite GEOS members have continuously performed long series of observations, no light curve has been published since these earlier studies remained unknown to the community[1]. However, some observers reported periodic variations in 1999 (Leonini, 2001).

Since 2011, KWS database provides annual short series of V and Ic measurements. Sjoerd Dufoer has performed good and long series of V and B CCD measurements for the GEOS since 2020; his data are available in the AAVSO database. We have also explored the data available in OMC-Catalogue optically variable source (Alfonso-Garzon et al., 2012).

Then, on the basis of these available data, it is now possible to discuss the long term light variations of V449 Cyg and its type as L or SR variable star.

## 2. Results from GEOS and AAVSO earlier visual observations

The Figure 1 shows the finding chart used by GEOS observers since the year 1974. The magnitude of the comparison stars have been measured by Michel Dumont (Dumont, 1983); they are in accordance with the magnitudes given in SIMBAD-CDS database in the Johnson UBV system (Høg et al., 2000).

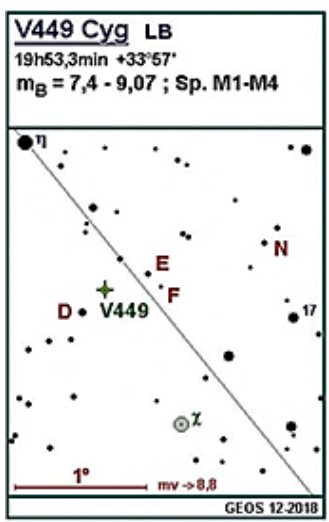

| Star | HD number | Magn. V GAIA (SIMBAD) | Magn. V (Dumont, 1983) |
|------|-----------|-----------------------|------------------------|
| D | 188484 | 6.76 | 6.79 |
| E | 226208 | 7.42 | 7.45 |
| N | 225806 | 7.86 | n/a |
| F | 226159 | 8.53 | 8.53 |

Figure 1 – V449 Cyg, GEOS finding chart and references for the comparison stars.

---

[1]. The *Notes Circulaires* of the GEOS are Now in open access (see the bibliography section).

Since the first years of the GEOS (1974-1975), the visual series on red stars have been processed following the GEOS ALCEP treatment program described in our 1987 study (Ralincourt, Ph., Poretti E., Boistel, G., 1987). Its principles are rather simple and described in the quoted paper: the mean magnitudes are computed for equal time-intervals (every 5 days in our case). For each observer, two parameters are calculated: the systematic difference in magnitude ($\Delta m$) and the standard deviation ($\sigma$). At the next iteration the new mean magnitudes are recalculated by affecting each individual measure with a corresponding weight coefficient equal to ($\frac{1}{10} \times \sigma^2$). Calculations are repeated until the final magnitudes remain stable.

From GEOS earlier observations (Figure 2, 1974; Figure 3, 1981-1982) it appears that V449 Cyg was mainly an irregular star of LB type (as defined by the GCVS – see section 5) showing irregular light variations with sometimes a pseudo-period of roughly 100 days (Figure 3).

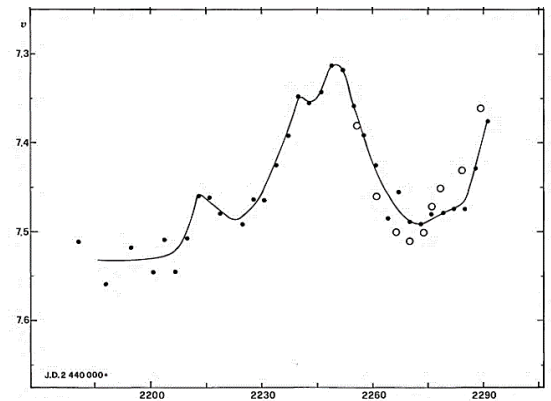

Figure 2 – V449 Cyg in 1974 (GEOS visual light curve with dots), completed with 9 photoelectric measurements (open circles) performed by Dr. L. Baldinelli for the GEOS (private communication). The good agreement between visual and photoelectric measurements justifies the ALCEP procedure[2].

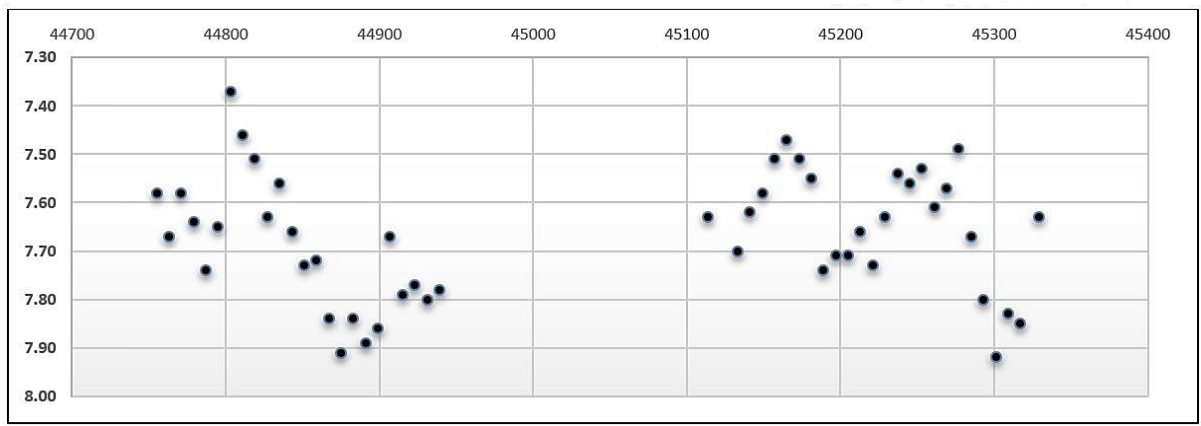

Figure 3 – V449 Cyg in 1981-1982, GEOS mean visual light curve processed by the ALCEP program; V449 Cyg shows irregular variations with sometimes a cycle of roughly 100 days (Boistel, 1987) (HJD+2400000).

We have searched for any material among old AAVSO visual observations with the help of AAVSO VSTAR software (Benn, 2012). The only useful material was found for the years 1955-1959 where several observers provided data on a long time-scale. Figure 4 show the light curves obtained for that period; there was a break in the data between the end of 1955 and the month of may 1956, so it was not possible to connect the two light curves. The figures 4a and 4b show the irregular variations of light for V449 Cyg from AAVSO mean light curves computed for this paper[3]. These light curves show some short cycles of 100 to 200 days and irregular variations between these cycles. Unfortunately we did not find any material in AAVSO database corresponding to the observations performed by the GEOS during the years 1974 or 1981-1982 in a perspective of comparison or pooling of data.

Since the end of the years 1990s, V449 Cyg seemed to become periodic, with a cycle of roughly 50 days, more in adequation with a SRB type, as Leonini (2001) showed with the help of 97 visual estimates (Figure 5), and as shown by the V-light curves obtained later with by KWS (Figure 6a and 6b).

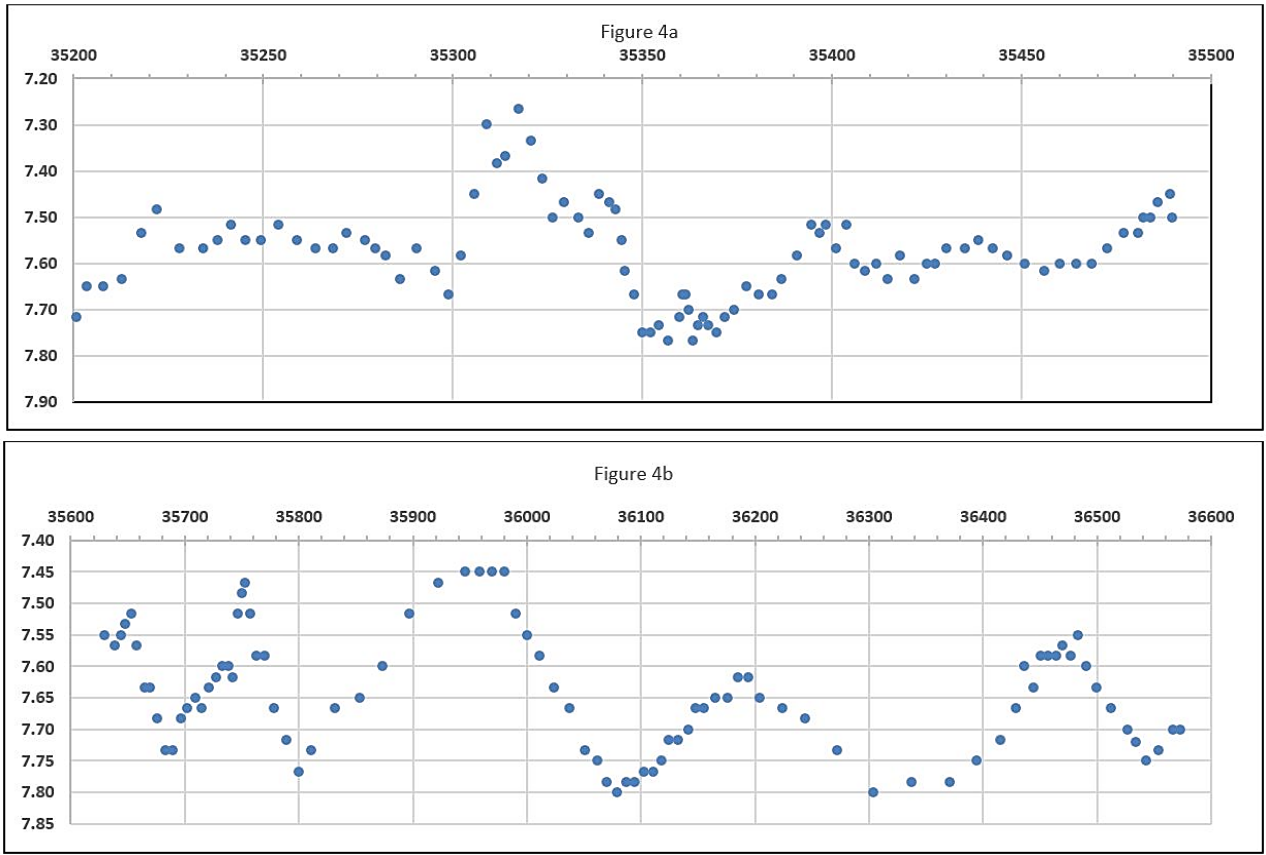

Figure 4a and 4b: V449 Cyg AAVSO old visual series (Not ALCEP processed). Fig. 4a: Mean light curve for the years 1955 (April)-1956 (January); fig. 4b: Mean light curve for the years 1956 (May)-1959 (February). V449 Cyg show irregular variations with sometimes 100 or 200 days pseudo-periods (HJD +2400000).

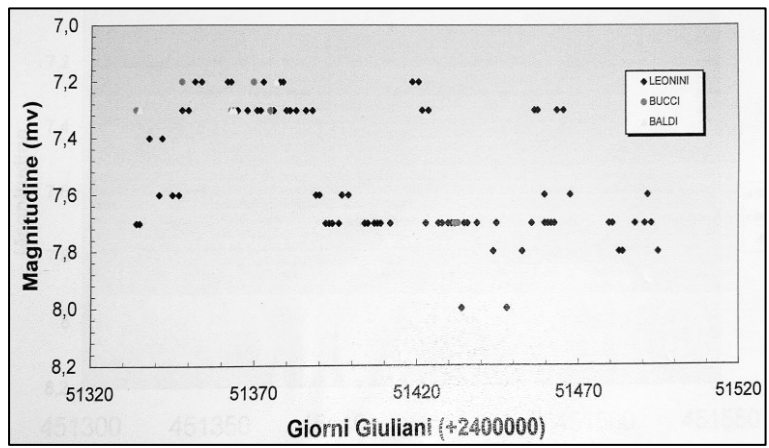

Figure 5 – Mean visual light curve of V449 Cyg in 1999 by italian observers – A short period of almost 50 days appears (Excerpt from Leonini, 2001).

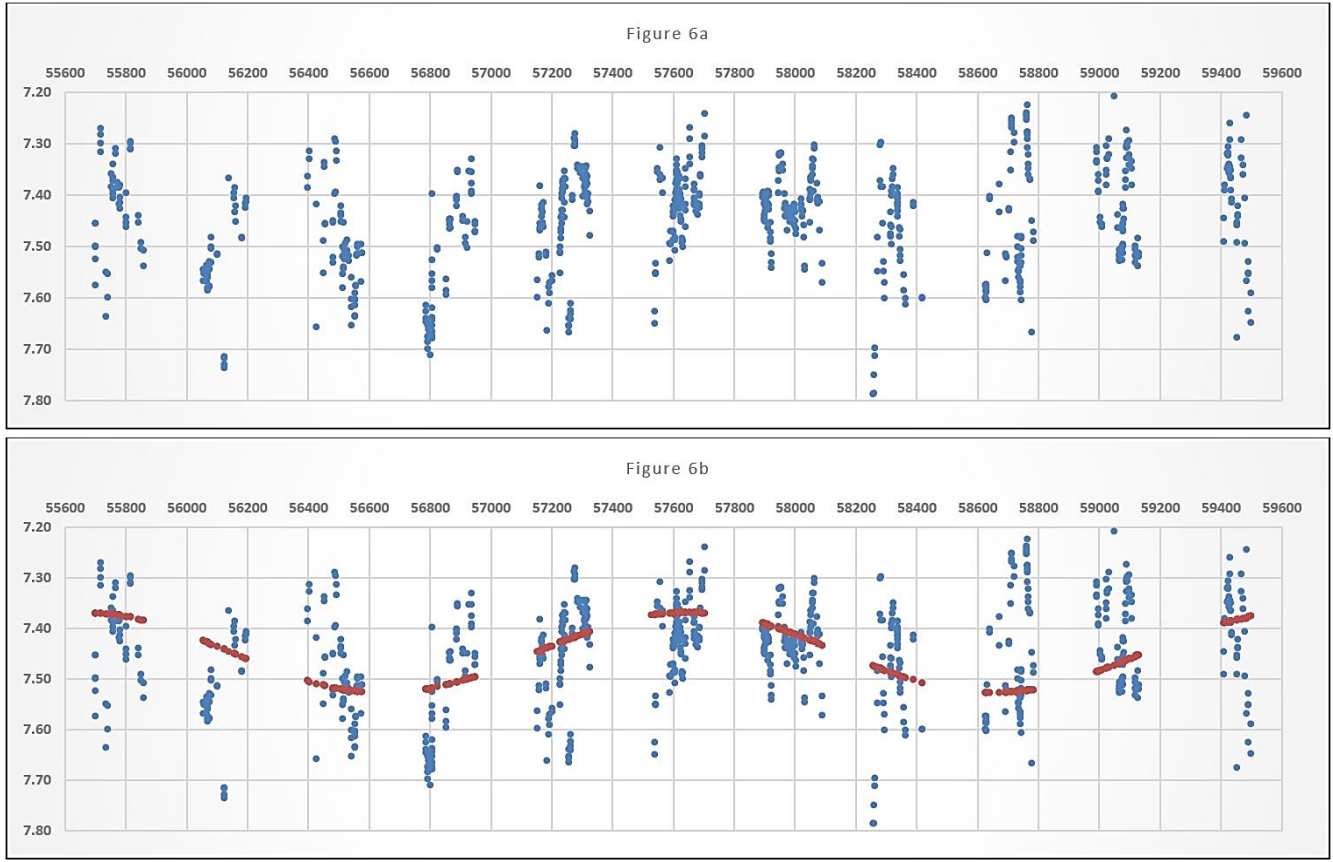

Figure 6: V449 Cyg. KWS V-series, 2011-2021. Fig. 6a, V light curve for the years 2011-2021. Fig. 6b - the same, with an empirical sinusoidal fit for a possible long term variation (of roughly 2000 days) (HJD+2400000).

### 3.  New observations: a study of 2020 GEOS visual observations and its comparisons with CCD-measurements performed by Jean-François Le Borgne and Sjoerd Dufoer

This section begins with a study and a comparison of 8 series of observations performed during the year 2020: 4 visual series performed by GEOS members (Mino Benucci – BEN, Guy Boistel – BTL, Stéphane Ferrand – FND, Jean-Claude Misson - MIS), and 4 CCD series performed by ASAS-SN (V filter), Jean-François Le Borgne - FLB (CCD unfiltered), Sjord Doefer – DFS (V filter) and KWS V-series (Table 1).

Table 1 – Time distribution for the 8 available series for the year 2020.

| V449 CYG 2020 | | JAN | FEB | MAR | APR | MAY | JUN | JULY | AUG | SEPT | OCT | NOV | DEC |
|---|---|---|---|---|---|---|---|---|---|---|---|---|---|
| VIS | BEN | | | | | | | | | | | | |
| | BTL | | | | | | | | | | | | |
| | FND | | | | | | | | | | | | |
| | MIS | | | | | | | | | | | | |
| CCD | FLB | | | | | | | | | | | | |
| | DFS | | | | | | | | | | | | |
| | ASASSN | | | | | | | | | | | | |
| | KWS | | | | | | | | | | | | |

Table 2 – Data details for the 8 series of observations of V449 Cyg analyzed for the year 2020 (HJD+2400000).

| Observer | Nb measures | HJD start | HJD end | Beginning | Ending | Nb days | Type |
|----------|-------------|-----------|---------|-----------|--------|---------|------|
| BEN | 149 | 58819 | 59178 | 01/12/2019 | 24/11/2020 | 359 | Vis. |
| BTL | 120 | 58895 | 59230 | 15/02/2020 | 15/01/2021 | 335 | Vis. |
| FND | 170 | 58865 | 59208 | 16/01/2020 | 24/12/2020 | 343 | Vis. |
| MIS | 90 | 58986 | 59160 | 16/05/2020 | 06/11/2020 | 174 | Vis. |
| ASAS-SN | 339 | 58904 | 59203 | 24/02/2020 | 19/12/2020 | 299 | CCD V |
| DFS | 143 | 59008 | 59334 | 07/06/2020 | 29/04/2021 | 326 | CCD V, B |
| FLB | 68 | 59000 | 59177 | 30/05/2020 | 23/11/2020 | 177 | Unfiltered CCD |
| KWS | 140 | 58991 | 59128 | 21/05/2020 | 05/10/2020 | 137 | CCD V |

Only one iteration of the ALCEP routine has been here necessary to correct the visual series from their standard deviation to compute a weighted mean as summarized in table 3. The figure 2 shows for example the good agreement between the 1974 visual light curve and the 9 photoelectric measurements (open circles) performed at that time by Dr. L. Baldinelli for the GEOS. The figure 7 shows the resultant mean visual light curve using the observations performed by the 4 GEOS members after the first iteration for the variations of V449 Cyg in 2020. Table 3 summarizes the statistical data from the processing of the 4 visual series.

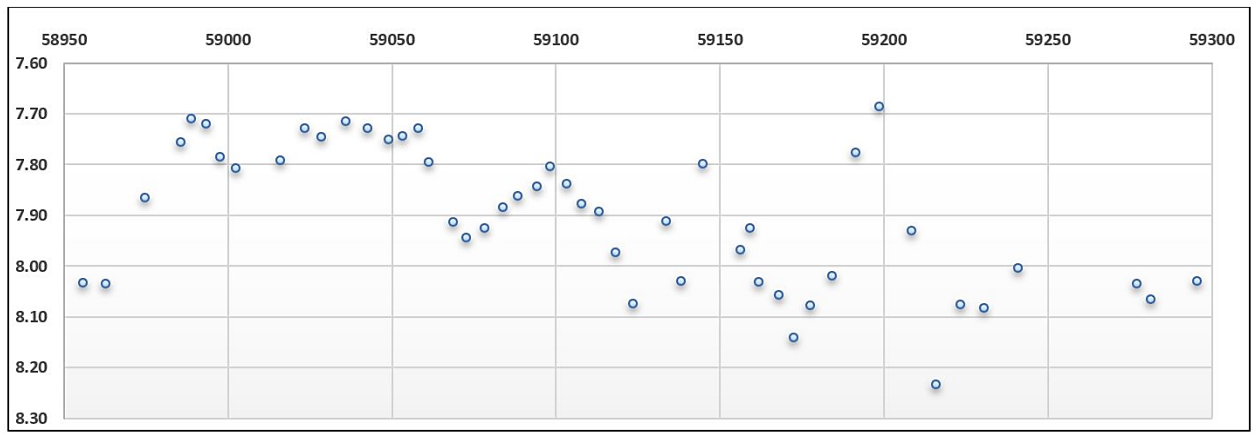

Figure 7 – V449 Cyg. GEOS visual mean light curve observations in 2020 (HJD+2400000).

Table 3 – Statistical data for the ALCEP processing of the 4 visual series (first iteration).

| Observer | Nb Measures | Systematic difference Δm (mag.) | Standard deviation σ (mag.) | Weight (1/10σ²) |
|----------|-------------|----------------------------------|------------------------------|-------------------|
| BEN | 149 | -0.011 | 0.115 | 7.62 |
| BTL | 120 | 0.023 | 0.136 | 5.40 |
| FND | 170 | 0.010 | 0.083 | 14.43 |
| MIS | 90 | -0.024 | 0.104 | 9.25 |

Figure 8 shows the light curves obtained by Jean-François Le Borgne (FLB) and Sjoerd Dufoer (DFS) with CCD measurements during the year 2020 over the same period. The difference between the two curves is easily explained by the fact that more red light from the red giant reaches the CCD device in the case of unfiltered CCD series than V filter because of the sensibility of the detector toward long wavelenght. At last, the figure 9 shows a comparison between the shifted GEOS visual mean curve (red dots), the DFS-CCD serie (blue dots) and the ASAS-SN g serie (green dots), the only useful set provided on this database. We have shifted the visual observations to the V ones to make a comparison as we did it previously in 1974 (Figure 2) and for the years 1981-1982 (Figure 3). A good agreement can be seen between these series. When the visual estimations are processed with the GEOS ALCEP program, the precision is much better than the 0.2 or 0.3 magnitude generally taken into account in various studies based on visual estimations extracted from AAVSO, BAA or AFOEV databases for example (Percy, Esteves et al., 2009). It is then possible to explore small amplitude light variations in red giants variations. As we can see on figures 8, 9 and 10, the visual maxima at HJD 2458935-940 and HJD 2459195 are confirmed by ASAS-SN light curve (Figure 9). But we can also see that ASAS-SN data are of lower quality than DFS' and FLB's ones. We then did not use these data sets for further period investigations.

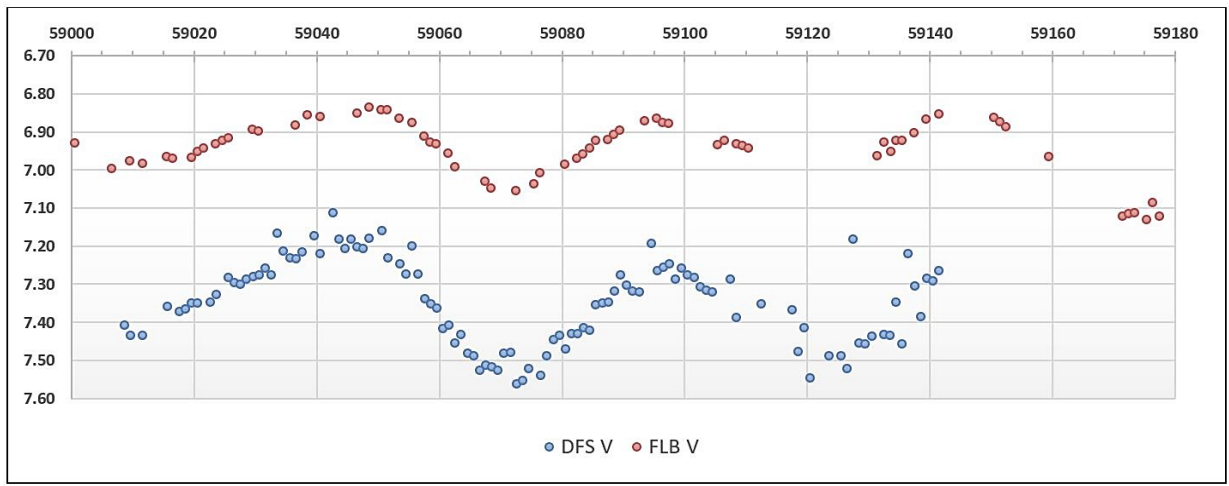

Figure 8 – V449 Cyg in 2020. DFS-V and FLB-CCD unfiltered measures (HJD+2400000).

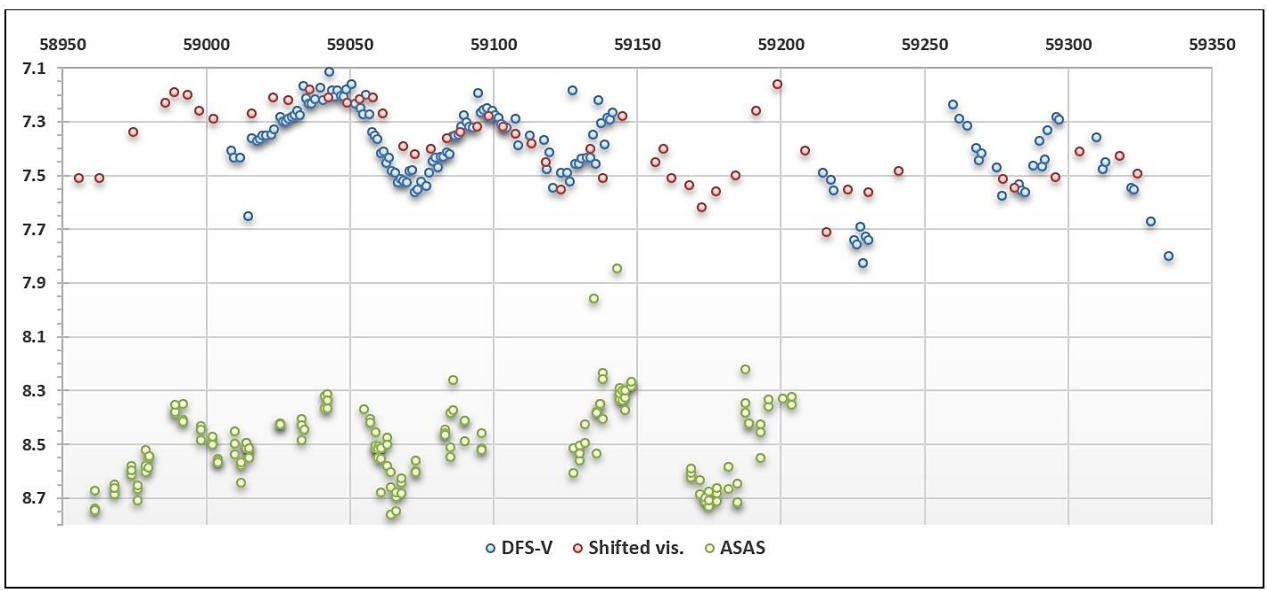

Figure 9 – V449 Cyg in 2020. Comparison between raw V-DFS measures (blue dots), shifted GEOS visual mean light curve (red dots) and raw ASAS-SN g-serie (green dots). The visual maxima at HJD ~58990 and ~59195 are confirmed by ASAS-SN measurements (HJD+2400000).

4. **Analysis of the period of V449 Cyg on the basis of the CCD-measurements obtained by Sjoerd Dufoer from 2020 to 2022, and other V-light variations obtained with automatic telescopes.**

Figure 10 shows the light curve of V449 Cyg obtained with the V measurements performed by Sjoerd Dufoer between 2020 and the beginning of 2022. This serie clearly shows a short cycle of 50-54 days with some irregularities.

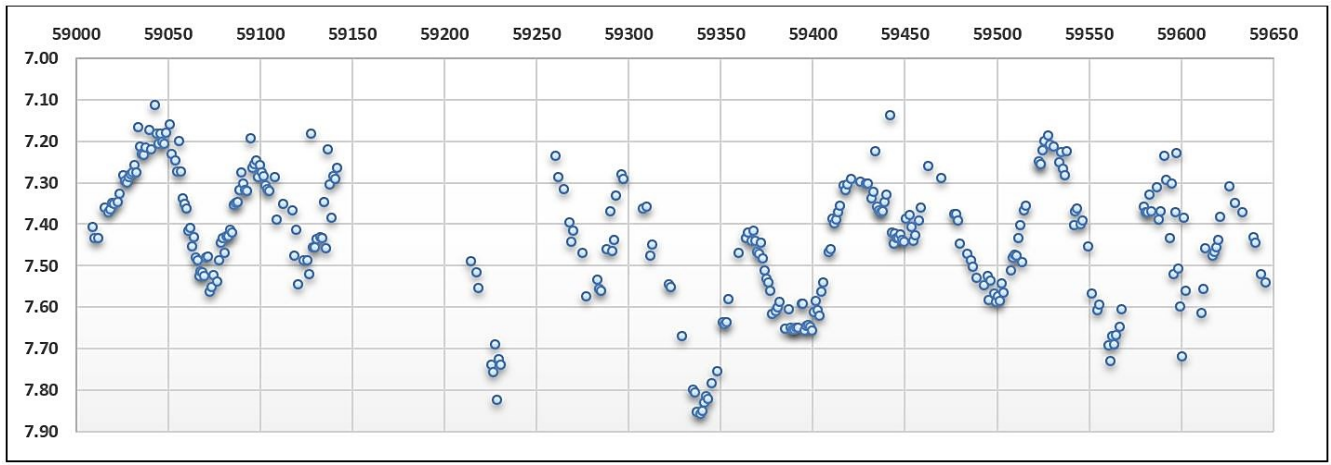

Figure 10 – V449 Cyg. V-DFS measures for 2020-2022 (HJD+2400000).

For a period analysis, we have first derived times of maximum for V449 Cyg from figures 8, 9 and 10. Table 4 gives mean times of maximum deduced from all these sets of observations on the 2019-2021 period (with help of custom polynomial fit provided by Peranso© software). The mean value for the period determined with these 11 times of maximum is 54.1 days, i.e., roughly 54 days, with a standard deviation of roughly 7 days. On a second hand, we used Peranso© software (Paunzen, E., Vanmunster, T., 2016) to run various routines for which we can control the parameters and the results: PDM (Stellingwerf), CLEANest (Foster), FERRAZ-MELLO ones. These routines give for the period of the variations of V449 Cyg: P ≈ 53.98 to 54.25 days. The Deeming FFT procedure gives a period of 54.44 days as illustrated on figures 11a to 11c). The Power Spectrum from AAVSO-VSTAR analysis tools gives a period of 54 days too (Figure 11d). All these results are very consistent with each other.

Table 4 – List of maxima of V449 Cyg during 2020-2022 extracted from figures 8 to 10.

| HJD 2400000+ | Visual | CCD | Difference (days) |
|---|---|---|---|
| 58985 | X | | |
| 59040 | X | X | 55 |
| 59095 | X | X | 55 |
| 59140 | X | X | 45 |
| 59195 | X | | 55 |
| 59250: | | X | 55 |
| 59295 | | X | 45 |
| 59360 | | X | 65 |
| 59415 | | X | 55 |
| 59460 | | X | 45 |
| 59525 | | X | 65 |
| 59580 ? | | X | 55 |
| | **Mean value →** | | **54.1** |

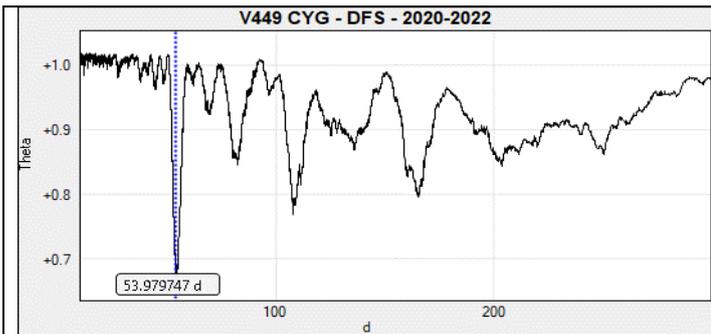

Fig. 11a – Period analysis: PDM routine (Peranso). P ~ 53.8 days.

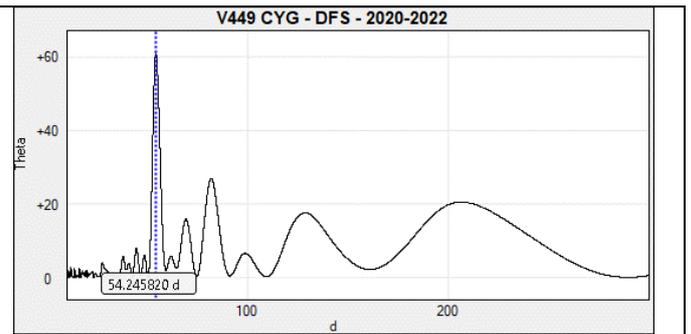

Fig. 11b – Period analysis: Ferraz-Mello routine (Peranso). P ~ 54.2 days.

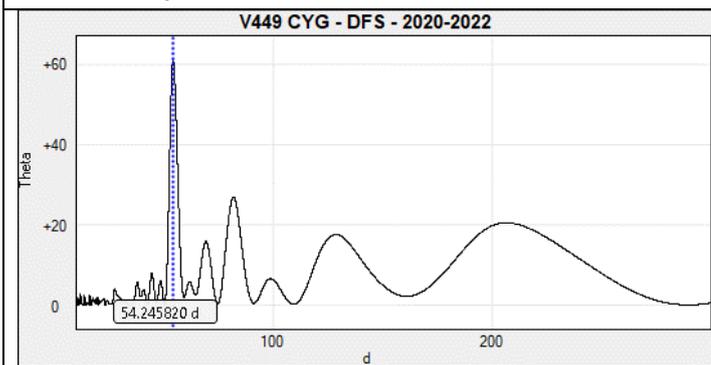

Fig. 11c – Period analysis: CLEANest routine (Peranso). P ~ 54.2 days.

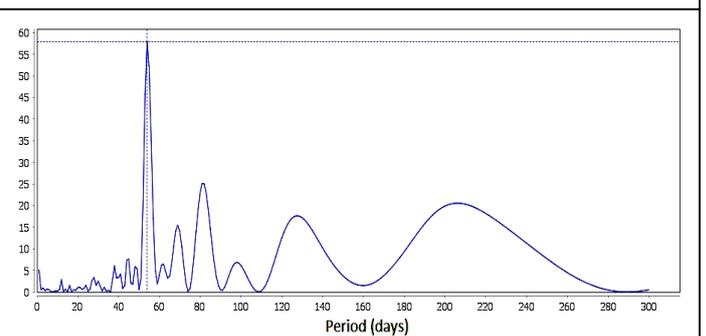

11d – Period analysis: Power Spectrum on DFS 2020-2022 V measurements (from AAVSO VSTAR analysis tools). P ~ 54 days.

Figure 11a-11d: Period analysis: various routines applied to DFS 2020-2021 V measurements (from Peranso© and AAVSO-VSTAR© softwares).

## 5. A long term survey of the variations of V449 Cyg with KWS time series, 2011-2021.

The figures 6a and 6b show the KWS long term series as given by KWS server; some obvious inaccurate measurements have been removed from the light curves. The KWS observations are made on a short duration of 5 months when visual observations are made on a full year, as CCD measurements obtained by DFS which cover a long-time scale. But we can clearly see the short period variations of 50 days (Fig. 6a). A longer and smooth variation of the mean magnitude appears and seems rather pseudo periodic (Fig. 6b). From an empirical sinusoidal fit we performed [$mag. = 7.45 + 0.08 \times sin(HJD \times 2\pi/2000 - 58.5)$], a long period of 2000±200 days cannot be excluded. Despite the fact that KWS annual series are short and uncomplete, we have applied a PDM routine to the whole set of observations provided by KWS (2011-2021). The periodogram is not significant, with numerous alias caused by the time distribution of the data (peaks around one year for example). This is expected as the observations time interval is not even equal to twice the suspected period. Then, we have focused on the years 2017-2020 for which the data are the most consistent ones. The same period search routines applied to KWS series (2018-2020) give periods from 54.5 to 54.57 days (Figure 12). The other peaks are still quite not significant; as said, they largely result from the time distribution of the observations taken into account, annually spread over only 4 to 5 months.

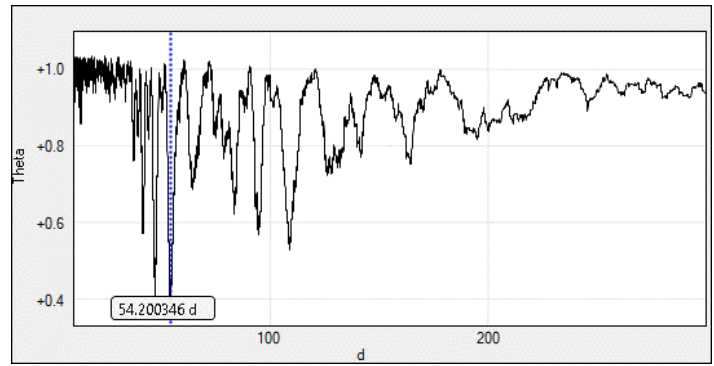

Figure 12: Period analysis: PDM routine applied to KWS 2018-2020 V-series (Peranso).

Figure 13 shows a phase plot on the DFS 2020-2022 observations (From the light curve shown in figure 10) computed on the period of 54.25 days; the dispersion is rather small for a semi-regular star.

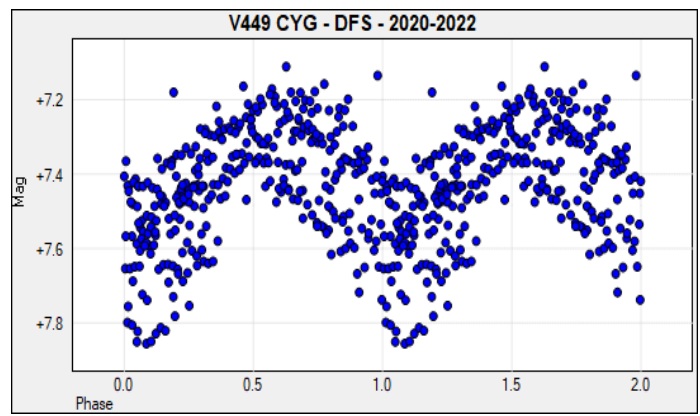

Figure 13: Phase plot for DFS 2020-2021 V-series, on the period 54.25 days (Peranso).

## 6. Conclusion

The SRB and L types are defined thus in the GCVS (Samus et al., 2017):

« **SRB** for **Semi-regular late-type** (M, C, S or Me, Ce, Se) giants with poorly defined periodicity (mean cycles in the range of 20 to 2300 days) or with alternating intervals of periodic and slow irregular changes, and even with light constancy intervals (RR CrB, AF Cyg). Every star of this type may usually be assigned a certain mean period (cycle), which is the value given in the Catalogue. In a number of cases, the simultaneous presence of two or more periods of light variation is observed.

**L** for **Slow irregular variables.** The light variations of these stars show no evidence of periodicity, or any periodicity present is very poorly defined and appears only occasionally. Like for the type I, stars are often attributed to this type because of being insufficiently studied. Many type L variables are really semiregulars or belong to other types. The **LB type** refers to slow irregular variables of late spectral types (K, M, C, S); as a rule, they are giants (CO Cyg). This type is also ascribed, in the GCVS, to slow red irregular variables in the case of unknown spectral types and luminosities. »

As we can see, the L type often encloses stars we don't have enough elements to classify with certainty. But the visual and CCD observations performed since the end of the 1990s and analyzed in this paper, show that V449 Cyg has entered on a new type of periodic variability, more in adequation with the SR type, perhaps the SRB type. Can we find, in the astrophysical data available, some elements to say about the kind of « cross-boundaries » that V449 Cyg reveals ?

Let us sum up the known elements for V449 Cyg in table 5. In a GAIA absolute G-band magnitude versus Median <BP – RP> diagram, V449 Cyg is among the long-period and SR variables, as shown by Gigoyan in its figure 4 with the coordinates {2.65; -2.26} (Gigoyan, 2020); we have reproduced this figure with the place of V449 Cyg in our figure 14. Its spectrum remains undetermined to tell more about the evolutionary path of V449 Cygni. Moreover, the IR data taken from the 2MASS database, put V449 Cyg {0.25; 0.91} on the M giants sequence (Bessell and Brett, 1988, fig.5).

Table 5: Astrophysical data for V449 Cyg (from VizieR, GAIA EDR3, CDS database). 2MASS IR sources data have been added. Data marked (*) are computed in this paper; they are consistent with GAIA EDR2 Catalogue and SIMBAD data (I/345/gaia2) and Anders et al. (2022) catalog (**): d = 0.496 kpc; absolute G mag. = -2.44 (and BP-RP = 2.55)

| Source GAIA DR2 and EDR3 | 2035185685071074944 |
|---|---|
| Spectrum | M1-M4 |
| Median V mag. | 7.42 |
| Median B Mag | 8.98 |
| Median B – V | 1.56 |
| Period (days) (*) | 54.2 |
| Log P (*) | 1.73 |
| G mag. | 6.235811 |
| BP mag | 7.701571 |
| RP mag | 5.044626 |
| BP - RP | 2.656945 |
| $T_{eff}$ (K) | 4000 |
| Log T | 3.6 (3.0 in VizieR) |
| Parallax (mas) (EDR3 Catalogue) | 1.77 – 1.97 |
| Distance d (pc) (*) | 500 – 540 (**) |
| Radial speed (km/s) | -140.03 |
| Absolute G mag. (*) | -2.26 (**) |
| 2MASS J mag. | 3.349 |
| 2MASS H mag | 2.439 |
| 2MASS K mag. | 2.193 |
| J-H mag. index | 0.91 |
| H-K mag. index | 0.25 |

Finally, from AAVSO visual series performed between 1955 and 1959, and GEOS visual 1970-1980s observations in good agreement with photoelectric measurements, V449 Cyg was mainly an irregular star with sometimes a 100 (even 200) days cycle, corresponding to L-type as GCVS defines it.

From the end of the years 1990, on the basis of the V-series performed by Sjoerd Dufoer and good available visual series, V449 Cyg is mostly periodic with a quasi-regular 54.2 ± 1 days cycle, in adequation with the SRB type (pseudo period with irregular changes). From the KWS long V-series and despite their relative dispersion, a long period of about 2000±200 days should not be excluded to describe the long term light variations of V449 Cygni.

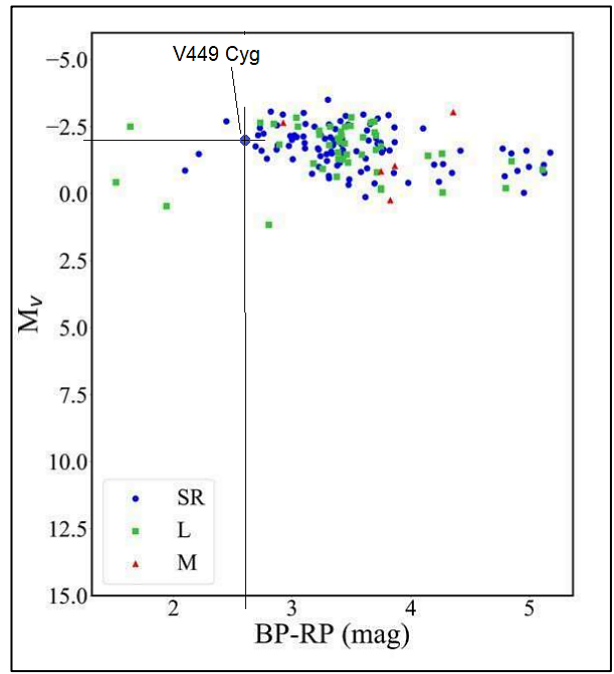

Figure 14: V449 Cyg in a GAIA absolute mag. vs <BP-RP> mag. diagram. Adapted from Gigoyan, 2020, Fig.4, p. 203. V449 Cyg {2.65; -2.25} is then among SR variables.

## 7. References

### Web sites and facilities for the processing of electronic data:

AAVSO VSX international server: https://www.aavso.org/vsx/

AAVSO VSTAR software: https://www.aavso.org/vstar-overview

AFOEV archives on CDS ftp: http://cdsarc.u-strasbg.fr/ftp/pub/afoev/cyg/v449

ASAS-SN Sky Patrol, variable stars server: https://asas-sn.osu.edu/variables

ASAS-SN PATROL international database: https://asas-sn.osu.edu (Shappee et al., 2014; Kochanek et al., 2017)

BAA-VSS Photometry Database: https://britastro.org/photdb/

GCVS catalogue: http://www.sai.msu.su/gcvs/gcvs/

GEOS Open Access Publications: http://geos.upv.es/index.php/publications

KWS (Kamogata/Kiso/Kyoto Wide-field Survey) international server: http://kws.cetus-net.org/~maehara/VSdata.py (Maehara H., 2014)

OMC-Catalogue optically variable source international database: http://cdsarc.u-strasbg.fr/ftp/J/A+A/548/A79/ (V449 Cyg = OMC2677000060).

PERANSO© (Light curve and PERiodANalysisSOftware): https://www.cbabelgium.com/peranso/

SIMBAD-CDS international database: http://simbad.u-strasbg.fr/simbad/

**Aknowledgements**: We want to thank Simone Leonini and Pietro Baruffetti (from Unione Astrofili Intaliani) for the communication of the important paper published in the review *Astronomia*.